\documentclass[aps,prl,twocolumn,preprintnumbers,amsmath,amssymb,nobibnotes]{revtex4}
\usepackage{graphicx}% Include figure files
\usepackage{bm}% bold math

\newcommand{\micron}{\ {\rm \mu m}}
\newcommand{\meV}{\ {\rm meV}}

\begin{document}

%\preprint{APS/123-QED}

\title{Polariton boxes in a tunable fiber cavity}

\author{Benjamin Besga}\thanks{These authors contributed equally to this work.}
\author{Cyril Vaneph}
\author{Jakob Reichel}
\author{J\'er\^ome Est\`eve}
\affiliation{Laboratoire Kastler Brossel, ENS/CNRS/UPMC, 24 rue
 Lhomond, 75005 Paris, France}
\author{Andreas Reinhard\textsuperscript{1}}\thanks{These authors contributed equally to this work.}
\author{Javier Miguel-S\'anchez\textsuperscript{1}}
\author{Ata\c{c} Imamo\u{g}lu\textsuperscript{1}}
\author{Thomas Volz\textsuperscript{1,2}}
\affiliation{\textsuperscript{1}Institute for Quantum Electronics,
ETH Zurich, 8093 Zurich, Switzerland \\ \textsuperscript{2}Centre
for Engineered Quantum Systems, Department of Physics and Astronomy,
Macquarie University, North Ryde, NSW 2109, Australia}
\date{\today}

\begin{abstract}

\end{abstract}

%\pacs{xxx}
\maketitle

{\bf Cavity-polaritons in semiconductor photonic structures have
emerged as a test bed for exploring non-equilibrium dynamics of
quantum fluids in an integrated solid-state device
setting~\cite{Deng:RevModPhys10}. Several recent experiments
demonstrated the potential of these systems for revealing quantum
many-body physics in driven-dissipative
systems~\cite{Carusotto:RMP13}. So far, all experiments have relied
on fully integrated devices with little to no flexibility for
modification of device properties. Here, we present a novel approach
for realizing confined cavity-polaritons, that enables in-situ
tuning of the cavity length and thereby of the polariton energy and
lifetime. Our setup is based on a versatile semi-integrated
low-temperature fiber-cavity platform~\cite{Colombe:Nature07,
Hunger:NJP10, Muller:APL10, Sanchez:NJP13}, which allows us to
demonstrate the formation of confined polaritons (or polariton
boxes) with unprecedented quality factors. At high pump powers, we
observe clear signatures of polariton lasing~\cite{Imamoglu:PRA96}.
In the strong-confinement limit, the fiber-cavity system could
enable the observation of the polariton-blockade
effect~\cite{Verger:PRB06}.}

\begin{figure}[t!] %  figure placement: here, top, bottom, or page
   \centering
   \includegraphics[width=86mm]{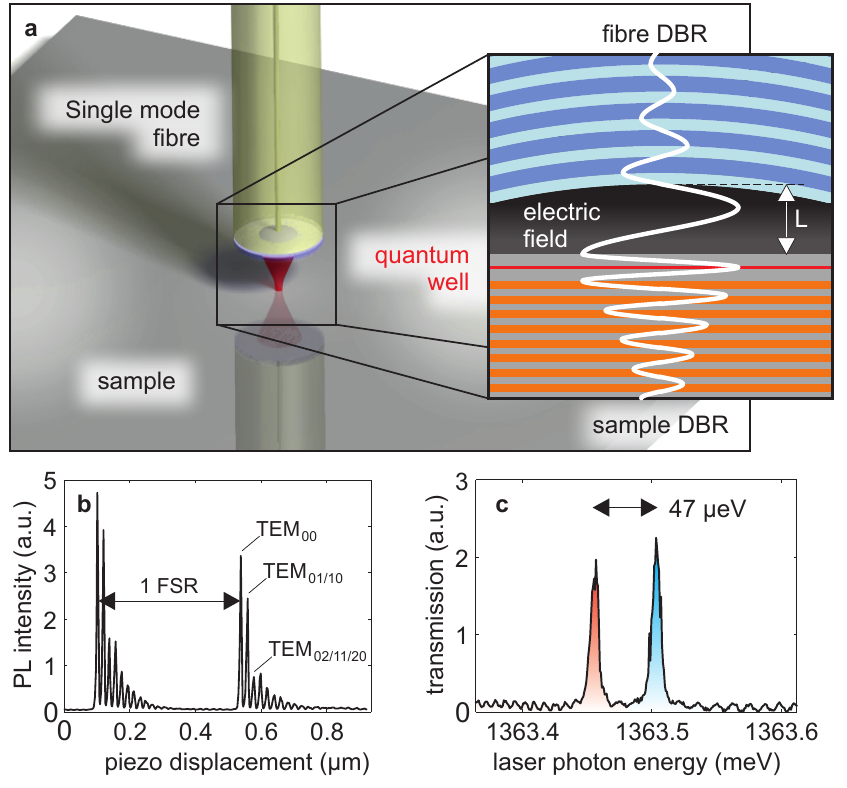}
   \caption{
{\bf Fiber cavity microscope. a}, Semi-integrated microcavity
consisting of a curved fiber-end mirror and an integrated planar DBR
mirror, with a quantum well grown on top. The inset sketches the
electric field inside the cavity. The quantum well is sandwiched
within the surface layer of the semiconductor DBR, in a way that the
electric field at its position is maximal. The cavity length $L$ is
defined as the distance between the two mirror surfaces and is
controlled by a piezo actuator. {\bf b}, Photoluminescence intensity
at a fixed photon energy (at a center wavelength of $909.2$~nm,
filtered with a spectrometer) far red-detuned from the QW resonance
as a function of piezo displacement (sample C). Fundamental modes
(TEM$_{00}$) are separated by a free spectral range (FSR) of
$\lambda/2$. Additional peaks correspond to groups of higher-order
transverse modes, e.g. TEM$_{01/10}$ or TEM$_{02/11/20}$. {\bf c},
Resonant transmission of the QW-cavity system (sample A) recorded on
a photodiode mounted underneath the sample. Here, the cavity length
is fixed while we sweep the wavelength of the resonant laser. With
this technique, we resolve a polarization splitting of a TEM$_{00}$
mode of $47~\mu\mathrm{eV}$, which is not visible in panel {\bf
b}.}
\end{figure}

Semiconductor microcavity polaritons are the elementary excitations
resulting from the strong coupling of quantum-well (QW) excitons to
photons confined in a high-Q microcavity~\cite{Weisbuch:PRL92}. Due
to their half-matter half-light character, cavity polaritons inherit
the effective mass of the photons and at the same time interact with
each other through the excitonic part of their wavefunction. In a
3D-confined polariton box, the average distance between two
polaritons determines the strength of polariton-polariton
interactions: The smaller the spatial extent of the polariton mode,
the larger the interaction strength between polaritons. While such
polariton boxes enable non-linear optical behavior, the
single-photon non-linear regime has not been reached. In contrast,
pronounced single-photon nonlinearities have been demonstrated with
semiconductor self-assembled quantum dots strongly coupled to a
cavity mode, where tight electronic confinement on nanometer length
scales leads to a highly anharmonic quantum
system~\cite{Faraon:NatPhys08, Reinhard:NatPhot12}. For QW
polaritons, confinement can be induced through post-growth
engineering of the photonic and/or the excitonic part of the
wavefunction. Exciton confinement has been achieved by controlled
application of sample stress~\cite{Nogoita:APL99, Balili:Science07},
naturally occurring defects~\cite{Kasprzak:Nature06}, and recently
by light-induced creation of spatially modulated excitonic
reservoirs~\cite{Wertz:NatPhys10,Amo:PRB10,Tosi:NatPhys12}. In
particular the latter method provides very precise control over the
potential landscape. However, the achievable confinement length
scales are limited by the excitonic diffusion length of
several$\micron$. In addition, the presence of the large exciton
reservoir and the additional off-resonant laser light might pose
strong limitations for resonant probing techniques. Polariton
confinement through its photonic part has been demonstrated by the
use of mesa structures~\cite{ElDaif:APL06},
micropillars~\cite{Ferrier:PRL11} and photonic-crystal
cavities~\cite{Azzini:APL11}. However, these systems typically
suffer from enhanced excitonic or photonic losses, once the spatial
dimensions approach the $\micron$-scale. In addition, their spectral
tunability is limited.

Here, we demonstrate a new method for the implementation of tight
polaritonic confinement which at the same time allows the
experimenter to tune in-situ both energy and lifetime of the
polaritons. Our semi-integrated cavity system consists of a concave
dielectric distributed Bragg reflector (DBR) deposited at the end of
an optical fiber tip, and a sample chip containing a semiconductor
DBR and an optically active quantum-well layer grown on top (see
Fig.~1a and Methods). The Fabry-Perot type cavity defines TEM
Gaussian modes and thereby imprints the resulting polaritonic
wavefunctions, yielding a lateral polariton confinement by
all-optical means. Tuning the length of the cavity allows both for
adjustment of the polariton lifetime and the precise tuning of the
polariton energy at one and the same spot of the sample area. Using
a single InGaAs quantum well, we observe an avoided level crossing
of the polariton spectral lines as a function of cavity length,
giving a Rabi splitting of 3.4~meV, consistent with previously
reported values in literature. For a sample containing 9 quantum
wells, we observe clear signatures of polariton lasing.

Figure 1a sketches the setup of our fiber-cavity system: The fiber
is mounted a few micrometers above the surface of the semiconductor
chip which consists of an AlAs/GaAs semiconductor DBR with the
active InGaAs QW layers grown on top. A 3D-stack of attocube
nanopositioner stages controls the position of the sample relative
to the fixed fiber tip. The system is kept at $4.2$~K by immersing
it in a liquid helium bath. For the experiments described here, we
used four different samples (A, B, C, D) with 1, 4 or 9 QWs (for
details see Methods). In order to probe the system, we first
performed photoluminescence (PL) spectroscopy with $780$~nm
excitation light. The fiber was used both for excitation and photon
collection. Many of the intrinsic properties of the cavity can be
inferred from PL spectra with the cavity mode far-detuned from the
exciton resonance. Figure~1b shows the recorded PL-intensity at a
fixed wavelength of 909.2~nm as a function of cavity length. In
addition to two fundamental transverse electromagnetic TEM$_{00}$
modes, which are separated by a free spectral range (FSR), the
diagram shows a multitude of equally spaced higher-order transverse
modes. All the cavity modes exhibit a polarization splitting that
typically is significantly larger than the cavity
linewidth~\cite{Hunger:NJP10} (see Figure~1c). The quality factor of
the cavity modes increases with the length $L$ of the gap between
the fiber tip and the substrate. It exceeds values of $2\ \times
10^{5}$ at lengths of around $L\approx 20 \micron$. A more relevant
figure of merit for a Fabry-Perot type cavity is the finesse which
we determined from lifetime measurements for a cavity completely
detuned from the exciton resonance. We found a finesse of up to
$13000$ which is presumably limited by the reflectivity of the
on-chip semiconductor mirror.
% <-- Ringdown data from 14.12.2011
Another key parameter of our system is the waist size of the
$\mathrm{TEM}_{00}$ cavity modes. We measured values as small as
$\omega_{0} \approx 2.5\micron$~\cite{Sanchez:NJP13} for a fiber
mirror with a curvature radius of about $R \approx 75\micron$. Based
on Gaussian beam optics, we estimate a minimal achievable waist size
of about $1.4\micron$ at a wavelength of $900$~nm (assuming
$R=10\micron$ and $L=3\micron$). This value for the confinement
length scale of polaritons is at least comparable to or even better
than currently feasible in mesas~\cite{ElDaif:APL06} or
micropillars~\cite{Ferrier:PRL11}.

\begin{figure}[t!] %  figure placement: here, top, bottom, or page
   \centering
   \includegraphics[width=86mm]{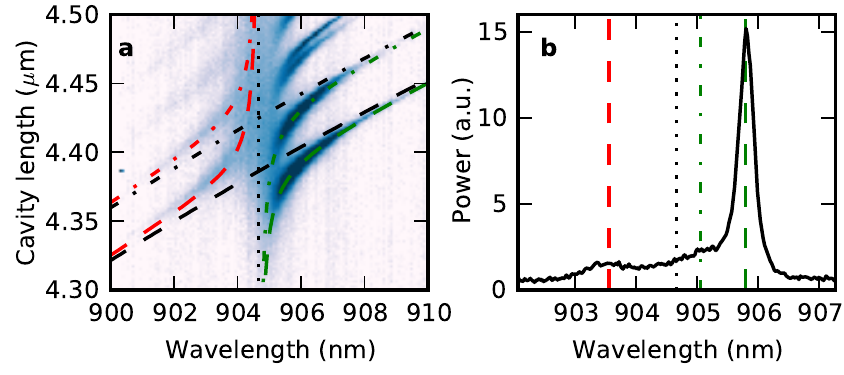}
   \caption{
\textbf{Avoided level crossing. a,} Spectrally resolved PL emission
under off-resonant excitation from the cavity containing a single
QW. A TEM$_{00}$ mode as well as higher-order transverse cavity
modes are tuned across the exciton resonance (black dotted line)
resulting in several pairs of UP and LP branches. The black dashed
(dashed-dotted) lines predict the spectral position of the uncoupled
TEM$_{00}$ cavity mode (TEM$_{01}$/TEM$_{10}$ modes). The red
(green) dashed lines are fits to the upper (lower) polariton
branches associated with the TEM$_{00}$ mode. Dashed-dotted lines
are fits to the polariton lines associated with the
TEM$_{01}$/TEM$_{10}$ modes. The fits yield a Rabi splitting of $2 g
\approx 3.4\meV$. {\bf b,} Cut through panel {\bf a} where the
TEM$_{00}$ cavity mode is resonant with the exciton line. The
TEM$_{00}$ lower-polariton peak (green dashed line), upper-polariton
peak (red dashed line) and the merged TEM$_{10}$/TEM$_{01}$
lower-polariton line (green dashed-dotted line) are visible.}
\end{figure}

In order to demonstrate the formation of polariton branches, we
approached the fiber tip close to the sample surface and performed
PL spectroscopy for varying cavity length. The resulting data for
sample D containing a single QW are shown in Figure 2a. The data
exhibit an avoided-level crossing for all the cavity modes crossing
the bare exciton line, clearly revealing the formation of
upper-polariton (UP) and lower-polariton (LP) branches. Note that
due to QW disorder, the LP resonance broadens and its amplitude
reduces significantly as it approaches the bare exciton resonance
(see Supplementary materials and Ref.~\cite{Savona:JPCM07}). All the
higher-order transverse modes couple to the QW excitons with
approximately the same strength as the fundamental TEM$_{00}$ mode,
which is theoretically expected due to the very weak dependence of
the coupling strength on the in-plane exciton momentum. A fit to the
first level crossings (green and red dashed and dashed-dotted lines)
yields a Rabi splitting of $2 g \approx 3.4$~meV. Figure 2b displays
a cut through the Fig.~2a at the cavity length where the fundamental
TEM$_{00}$ mode is resonant with the exciton line. While both UP and
LP branches are visible, the UP signal is much weaker and broader
due to fast non-radiative decay. Note that we confirmed the
formation of polariton branches by also probing their density of
states directly with a resonant tunable laser (see Supplementary
Material).

\begin{figure}[t!] %  figure placement: here, top, bottom, or page
   \centering
   \includegraphics[width=86mm]{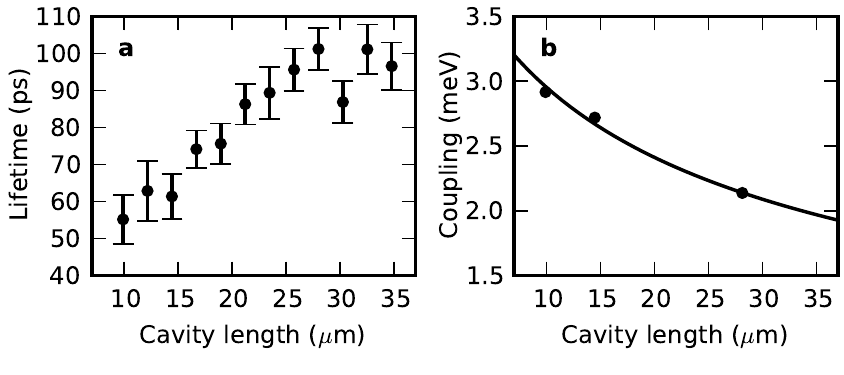}
   \caption{
{\bf Tuning the lower-polariton lifetime.} {\bf a,} Lifetime of the
LP resonance for a 4-QW sample as a function of cavity length. Each
point was taken with the LP resonance at the same wavelength,
$-5.1$~meV from the bare exciton resonance, i.e. different lengths
correspond to different longitudinal cavity-mode numbers. Up to a
length of $28\micron$, the lifetime increases linearly. The error
bars are $95~\%$ confidence bounds for the lifetime fits. {\bf b,}
With increasing cavity length $L$, the coupling strength $g$ between
exciton and cavity mode decreases as $g \propto 1/\sqrt{L+L_{\rm
DBR}}$. Using that expression together with the measured $g(L)$, we
calculate the LP exciton content in {\bf a} to vary between 40\% and
13\%.}
\end{figure}

The key feature of our system is its tunability. It enables in-situ
control of the cavity-exciton detuning, making the observation of an
avoided-level crossing in one and the same spot on the sample
possible. More importantly, when modifying the cavity length by
several$\micron$, the polariton lifetime can be significantly
prolonged.
%while only moderately reducing the coupling strength $g$. <-- T increases by a factor of 1.8, g decreases by a factor of 1.5, according to the figure.
Figure 3 illustrates this behavior for a sample containing 4 QWs
(sample B). The lifetime data in Figure 3a were recorded in
resonant-transmission spectroscopy using sub-picosecond laser pulses
with a bandwith of $10$~nm centered at the QW resonance (see
Supplementary Material). The results demonstrate ultra-long
polariton lifetimes of up to $100$~ps for an overall cavity length
of $28\micron$. For longer cavities, the measured lifetime
saturates. This could be due to diffraction losses at the fiber
mirror. %or increased influence of the uncoupled QW excitons for decreased coupling strength.
The coupling strength $g$ as a function of length (Fig.~3b) was
extracted from PL data similar to the ones of Figure 2a. The black
line in Fig.~3b is a fit to the data based on the assumption that
$g$ is proportional to the square root of the inverse effective
cavity length $L+L_{\mathrm{DBR}}$, where $L_{\rm DBR}$ is the
weighted penetration depth of the cavity field into the DBRs. It is
interesting to note that even though the physical length of our
microcavity is much longer than the one of a monolithic
semiconductor $\lambda$-cavity, for a length $L\approx 3.5\,\mu{\rm
m}$ the coupling strength $g$ reaches a value almost as high as in a
$\lambda$-cavity (see Supplementary Material).

\begin{figure}[h!] %  figure placement: here, top, bottom, or page
   \centering
   \includegraphics[width=\columnwidth]{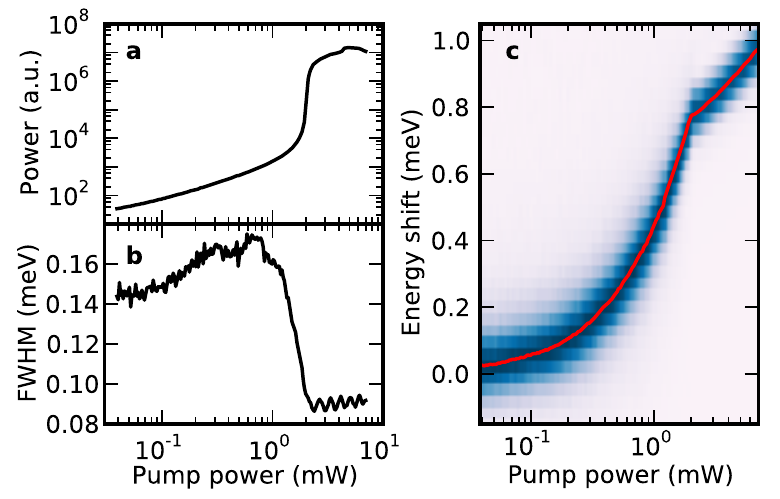}
   \caption{{\bf Polariton Lasing.} {\bf a}, The
   intensity of the light emitted from the lowest polariton mode
   LP$_{00}$ shows a highly non-linear behavior as a function of pump
   power with an increase by almost three orders of magnitude just above threshold.
   {\bf b}, The width of the LP$_{00}$ mode decreases sharply around
   threshold while the mode energy (panel {\bf c}) displays
   a continuous blueshift over the whole range of pump powers with
   distinctly different behavior above and below threshold.
   }
   \label{fig:setup}
\end{figure}

The long lifetimes reported here elevate our system into a new
domain that should enable the study of equilibrated long-lived
bosonic quantum fluids in an all-solid-state setting. In a first
experiment highlighting this potential, we demonstrate polariton
lasing in a 9-QW sample (C) consisting of three groups of three QWs,
each positioned at adjacent anti-nodes of the intra-cavity field.
Figure 4 displays the emitted PL intensity of the LP$_{00}$ mode as
a function of off-resonant excitation power for a very short cavity,
where the fiber almost touches the sample. We choose a
cavity-exciton detuning of $-5.8\,{\rm meV}\approx-1.3\,g$
corresponding to an exciton content of about 22\% at low pump power.
As is illustrated in Figure 4a), we observe a clear threshold
behavior with a sharp non-linear increase in the emitted number of
photons at around 2~mW of pump power. In addition, the linewidth of
the LP$_{00}$ mode starts to decrease and quickly drops below the
spectrometer resolution (Figure 4b) while the energy shift versus
pump power exhibits a kink. The latter behavior with the logarithmic
blueshift above threshold can be attributed to polariton lasing
\cite{Bajoni:PRL08,Roumpos:PRL10}. While the data taken here are
restricted to the more photon-like part of the LP branch, a suitably
designed sample with less exciton disorder, will allow the creation
of truly equilibrated long-lived polariton gases having a high
exciton content. Moreover, the tunability of our setup can be
explored to study the physics of polariton lasing for a broad range
of lifetimes and detunings. This should enable us to help
understanding open questions regarding the transition from polariton
to photon lasing \cite{Yamamguchi:PRL13, Ishida:arXiv13}.

In conclusion, we anticipate many exciting directions of research
with the present fiber-cavity setup. A further improvement in the
cavity-mode confinement would result in stronger interactions
between single polaritons. Assuming a typical polariton-polariton
interaction strength reported in the
literature~\cite{Amo:NatPhys09}, the regime of polariton blockade
would be within reach~\cite{Verger:PRB06}. In addition, the
highly-tunable fiber-cavity system lends itself to the observation
of the Feshbach-blockade effect proposed in
Ref.~\cite{Carusotto:EPL10}. More generally, our system opens up new
horizons for the study of strongly correlated states in polariton
fluids~\cite{Carusotto:PRL09}, including fractional quantum Hall
states of light~\cite{Ucucalilar:PRL12}. On a more practical note,
our semi-integrated platform combined with electrical injection of
carriers~\cite{Schneider:Nature13} could serve either as a
fiber-coupled tunable low-threshold polariton laser or a quantum
light source in the regime of polariton blockade.

This work is supported by NCCR Quantum Science and Technology
(QSIT), research instrument of the Swiss National Science Foundation
(SNSF), and an ERC Advanced Investigator Grant (A.I.). T.V.
acknowledges support through the Center of Excellence for Engineered
Quantum Systems, research instrument of the Australian Research
Council (ARC). The authors thank Iacopo Carusotto and Vincenzo
Savona for helpful discussions. The authors declare that they have
no competing financial interests. Correspondence and requests for
materials should be addressed to J.E. and T.V. (E-mail:
esteve@lkb.ens.fr, thomas.volz@mq.edu.au)

\section*{Methods}

\subsection*{Fiber and Samples}

Key element of our new microcavity setup (displayed in Figure 1a) is
the fiber mirror. In a first step, a concave mirror substrate in the
form of an indent is fabricated by ablating and surface-melting the
tip of an optical fiber using a CO$_2$-laser. In a subsequent step,
the fiber tip is coated with a dielectric mirror~\cite{Colombe:Nature07,Hunger:NJP10,
Sanchez:NJP13} (ATFilms, Boulder, Colorado). In the experiments, we
mostly used single mode fibers. Multimode fibers do not allow for
resonant reflection measurements but are equally suited for PL
measurements. In Figure 1b, a multi-mode fiber was used - in this
case more higher-order cavity modes couple to the fiber and are
therefore visible in the PL spectrum.

The sample mirror consists of several pairs of AlAs/GaAs
quarter-wave layers grown on a GaAs substrate by molecular beam
epitaxy. The InGaAs quantum wells were grown on top and sandwiched
between two GaAs spacer layers optimized for maximum coupling of QW
excitons to the cavity modes. For the experiments described in this
work, we used the following samples

\begin{table}[ht]
\caption{Samples} % title of Table
\centering % used for centering table
\begin{tabular}{c c c c c} % centered columns (5 columns)
\hline\hline %inserts double horizontal lines
Sample & \# QWs  & In content [\%] & QW thickness [nm] & \# DBR pairs  \\ [0.5ex] % inserts table
%heading
\hline % inserts single horizontal line
A & 1 & 8 & 10 & 30\\ % inserting body of the table
B & 4 & 8 & 11 & 28\\
C & 9 & 8 & 11 & 28\\
D & 1 & 8 & 11 & 28\\ [1ex] % [1ex] adds vertical space
\hline %inserts single line
\end{tabular}
\label{table:samples} % is used to refer this table in the text
\end{table}

\subsection{Polariton lifetime measurements}

For measuring polariton lifetimes, we excited the cavity through the
fiber using a pulsed white-light source and a band-pass filter,
resulting in a continuous spectrum of about $10~\mathrm{nm}$
bandwidth. The transmitted light of a TEM$_{00}$ LP mode was guided
out of the cryostat in a free-space configuration and sent to a
spectrometer for analysis. We then sent the light to an avalanche
photodiode (APD) in the Geiger mode and measured the ring-down time
of the polariton emission (see Supplementary Material).

\section{Supplementary Material}
\renewcommand{\thefigure}{S\arabic{figure}}
\setcounter{figure}{0}

\subsection{Role of exciton disorder}
From the PL spectra shown in Figure 2, we observe that the LP mode
broadens when the cavity mode approaches the QW-exciton resonance.
We attribute this effect to the presence of structural disorder in
the QW that is mostly due to alloy-fluctuations associated with the
high In content. Prior to inserting the sample into the fiber
cavity, we measured the QW-exciton free-luminescence spectrum and
observed a full-width at half-maximum of $1.4\meV$. To confirm that
this inhomogeneous broadening of the exciton line is responsible for
the observed PL spectra, we performed resonant-excitation
measurements for which quantitative predictions can be made more
easily. The transmission spectrum of the cavity mode is measured
using a photodiode glued to the backside of the sample, while tuning
the frequency of an excitation laser coupled to the fiber. We
repeated the experiment for different cavity lengths and fitted each
lower-polariton resonance to a Lorentzian. The curves in Figure~S1
show the dependence of the resonance widths and amplitudes as a
function of the peak central energy. In addition, we calculated the
expected transmission spectra using the model of an inhomogeneously
broadened ensemble of harmonic oscillators coupled to a cavity
mode~\cite{Houdre:1996iv,Diniz:2011dp,Kurucz:2011gu}. This model
describes excitons moving in a random potential, neglecting the
kinetic energy of the exciton~\cite{Whittaker:1998bb}.

The three parameters entering the model are the light-matter
coupling constant $g=1.65\meV$ that we extracted from the Rabi
splitting in PL measurements, the cavity linewidth $\kappa=18\,
\mu{\rm eV}$ that we extracted from the cavity transmission spectrum
far detuned from the QW, and the rms width $\Delta$ of the Gaussian
probability distribution describing the inhomogeneous ensemble of
oscillators. A fit to the data leads to $\Delta=1.1\meV$, which is
compatible with the observed free PL linewidth. The model (solid
lines in Figure~S1) quantitatively reproduces the increase of the LP
linewidth together with the decrease of the transmitted power when
the cavity mode is brought into resonance with the QW. The small
ratio $g/\Delta \approx 1.5$ results in disorder playing an
important role in the system dynamics. Increasing this ratio to a
value larger than 4, would render the presence of QW disorder almost
negligible~\cite{Houdre:1996iv,Diniz:2011dp,Kurucz:2011gu}.

\begin{figure}[htbp] %  figure placement: here, top, bottom, or page
   \centering
   \includegraphics[width=86mm]{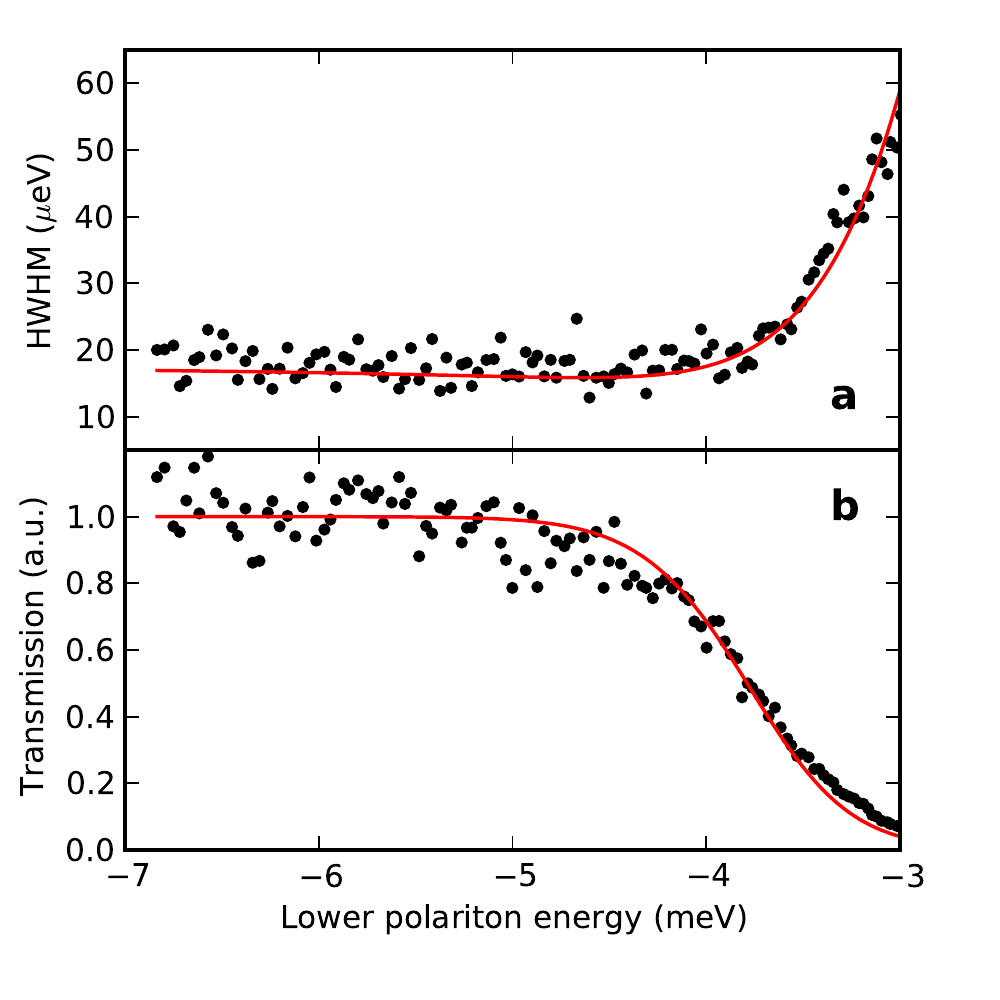}
   \caption{{\bf Characterization of the lower-polariton
   resonance via transmission measurements.} For a set of cavity
   lengths (within a single FSR), we measured the resonant transmission
   spectrum through the cavity using a tunable laser and fitted the
   observed lower-polariton resonance to a Lorentzian for each cavity length.
   {\bf a}, Half width at half maximum (HWHM) of the resonance as a
   function of the extracted polariton center energy, displayed in
   terms of the detuning with respect to the QW resonance.
   The detuning should be compared to the QW-cavity coupling strength $g=1.65\meV$.
   {\bf b}, Transmitted peak power as a function of its energy.
   At large negative cavity detuning, the observed width recovers the
   previously measured bare cavity-mode linewidth. Approaching the
   exciton resonance, the polariton broadens significantly. At the same time,
   the transmitted power drops. The solid lines show the result of a
   simplified model where the disordered quantum well is modeled by an
   ensemble of harmonic oscillators distributed with a $1.1\meV$ rms width
   Gaussian probability density.
   }
\end{figure}

\subsection{Quantum well--cavity coupling}
The coupling strength $g$ of a single QW to a micro-cavity depends
on two parameters: the oscillator strength per surface unit of the
QW that we denote by $f$ and the effective length $L_{\rm eff}$ that
characterizes the confinement of the cavity mode in the direction
perpendicular to the QW layer. The expression of $g$
is\,\cite{Panzarini:1999dh}
\begin{equation}
g = \left( \frac{\hbar^2}{4 \pi \epsilon_0} \frac{2 \pi e^2 f }{m \, L_{\rm eff}} \right)^{1/2},
\end{equation}
where $\epsilon_0$ is the vacuum permittivity, $e$ the charge
quantum and $m$ the electron mass. For a 2D planar micro-cavity, the
effective length can be defined as
\begin{equation}
L_{\rm eff} = \frac{\int 2 \, n^2(z) \, |E(z)|^2 \, dz}{|E(z_{\rm QW})|^2},\label{eq.Leff}
\end{equation}
where $n(z)$ ($E(z)$) gives the dependence of the refractive index
(electric field) along the cavity axis. $z_{\rm QW}$ denotes the
position of the QW.  For a microcavity where the mode is also
confined in the transverse $x,y$ directions, such as our fiber
Fabry-Perot (FFP) cavity, equation (\ref{eq.Leff}) still holds if the
paraxial approximation can be used to describe the cavity mode. In this
case $E(z)$ is the longitudinal dependence of the electric field,
which is the same as for a plane wave. Here, we also assume that the
transverse mode confinement is much larger than the exciton Bohr
radius.

\begin{figure}[h!]
  \centering
     \includegraphics[width=86mm]{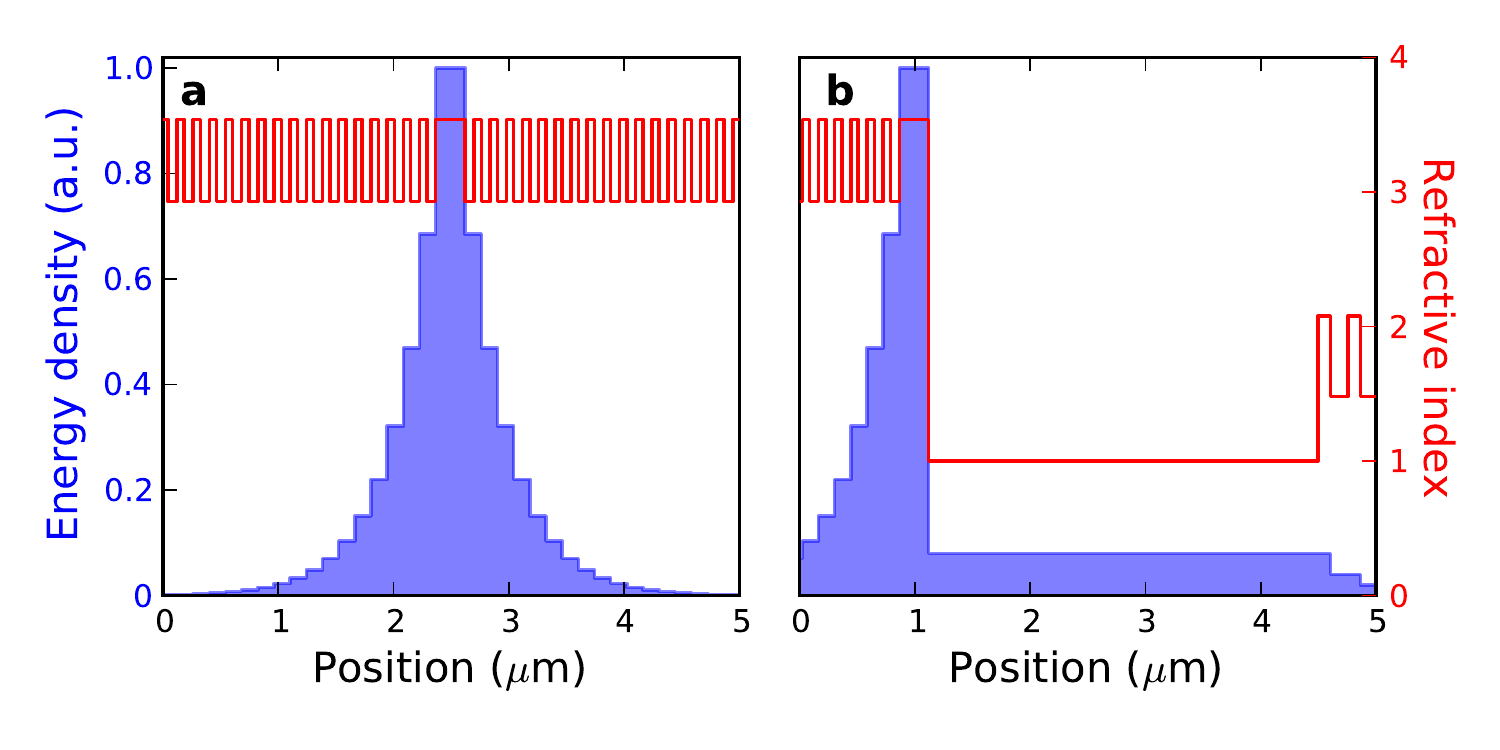}
  \caption{{\bf Comparison of the electric field confinement in a GaAs
  $\lambda$-cavity and in a FFP cavity.} Figure {\bf a} corresponds to
  a GaAs $\lambda$-cavity with AlAs/GaAs DBRs. Figure {\bf b}
  corresponds to a FFP cavity with a $L=3.5\,\mu$m long gap between the
  top dielectric mirror and the semiconductor sample surface.
  The red curves show the refractive index profile $n(z)$. The blue curves
  show the computed dependence of $n^2(z) |E(z)|^2$ using the transfer matrix method.
  The integral under the blue curve is (almost) the same in {\bf a} and {\bf b}, resulting
  in the same effective cavity length. This means equal QW-cavity coupling, even
  though the physical length of the FFP cavity is significantly longer than the
  one of the $\lambda$-GaAs cavity.}
\end{figure}

In a typical 2D planar GaAs microcavity, the QW is grown in the
middle of a GaAs layer with optical thickness $\lambda$. This
so-called $\lambda$-cavity is surrounded by two AlAs/GaAs DBRs (see
figure~S2). The electric field penetrates each DBR on a length scale
given by\,\cite{Panzarini:1999dh}
\begin{equation}
L^{\rm DBR}_{\rm AlAs/GaAs}= \frac{\lambda}{4} \frac{n_{\rm
GaAs}n_{\rm AlAs}}{n_{\rm GaAs}^2 (n_{\rm GaAs}-n_{\rm AlAs})}
\approx 0.31\,\mu{\rm m}, \label{eq.LDBR_GaAsAlAs}
\end{equation}
where we used $n_{\rm GaAs}=3.54$, $n_{\rm AlAs}=2.93$ and
$\lambda=900$\,nm. The effective length $L_{\rm eff}$ is the sum of
the optical length ($\lambda/n_{\rm GaAs}$) of the spacer layer plus
the penetration length of each mirror, multiplied by the square of
the refractive index of the spacer layer, here GaAs:
\begin{equation}
L_{\rm eff} = n^2_{\rm GaAs} \left( L^{\rm DBR}_{\rm AlAs/GaAs} +
\frac{\lambda}{n_{\rm GaAs}} + L^{\rm DBR}_{\rm AlAs/GaAs}\right).
\end{equation}
Inserting numerical values gives an effective length of $L_{\rm eff}
\approx 10.8 \,\mu{\rm m}$. Together with a typical oscillator
strength of $f=5 \times 10^{12}\,{\rm cm}^{-2}$\,\cite{Panzarini:1999dh},
this leads to a value of $g=1.8$\,meV.

In the case of our FFP cavity, the bottom half of the cavity has the
same structure as the $\lambda$-cavity described above. The QW is at
the center of a $\lambda/n_{\rm GaAs}$ thick GaAs layer grown on top
of an AlAs/GaAs DBR. The top DBR is made of dielectric materials
(Ta$_2$O$_5$/SiO$_2$) and is separated from the sample surface by a
gap (vacuum) of length $L$ (see Figure S2). The penetration length
in the top mirror is\,\cite{Panzarini:1999dh}
\begin{equation}
L^{\rm DBR}_{{\rm Ta}_2{\rm O}_5\rm{/SiO}_2}=
\frac{\lambda}{4}\frac{1}{(n_{{\rm Ta}_2{\rm O}_5}-n_{{\rm SiO}_2})}
\approx 0.38\,\mu{\rm m},
\end{equation}
where we use $n_{{\rm Ta}_2{\rm O}_5}=2.078$,
$n_{\rm{SiO}_2}=1.479$. This expression is different from
(\ref{eq.LDBR_GaAsAlAs}), because the dielectric DBR starts with a
high index layer (Ta$_2$O$_5$), whereas the AlAs/GaAs mirror starts
with a low index layer (AlAs). Taking into account that the lower
DBR mirror is surrounded by GaAs and the top one by vacuum, we
obtain the following expression for the effective length of the FFP
cavity
\begin{equation}
L_{\rm eff} = n_{\rm GaAs}^2 \left( L^{\rm DBR}_{AlAs/GaAs}
+\frac{\lambda}{n_{\rm GaAs}}\right) + L + L^{\rm DBR}_{{\rm
Ta}_2{\rm O}_5\rm{/SiO}_2},
\end{equation}
which gives the approximate expression
\begin{equation}
L_{\rm eff} \approx L + 7.45 \,\mu{\rm m}
\end{equation}
Despite its much longer physical length, a FFP cavity with a gap of
3.5\,$\mu$m therefore has the same effective length than a
monolithic GaAs $\lambda$-cavity.

\subsection{Measurement of polariton lifetimes}

%Using a resonant light source, system transitions are directly probed and complicated relaxation mechanisms occurring in off-resonant excitation schemes can be avoided. To this end, we coupled the light into the cavity via the fiber port. When not resonant with a cavity (or polariton) mode, the light is simply back-reflected from the fiber mirror and does not interact with the microcavity system. Only light that is close to a cavity (polariton) resonance is injected into the cavity and mainly leaks out of the system through the semiconductor (sample) mirror after a short delay corresponding to the lifetime of the cavity photon (polariton).

\begin{figure}[t!]
  \centering
     \includegraphics[width=86mm]{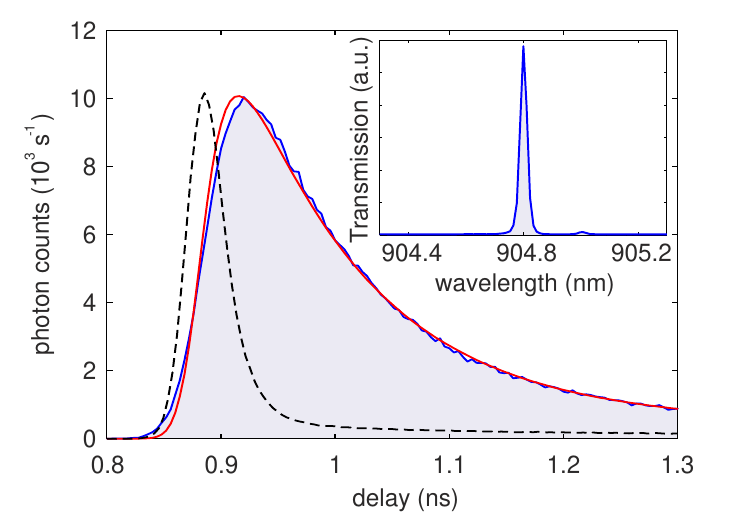}
  \caption{{\bf Resonant ring-down measurements.} Ring-down trace
  recorded on the four-QW sample. The data (blue solid line) was
  recorded at a cavity length of about $28~\mu\mathrm{m}$. The dashed
  line shows the time response when directing the broad-band light
  directly to the APD and thus represents the time-response function
  of the APD. The red line shows the convolution of the APD-response
  function with the best-fit exponential decay with time constant
  $(101 \pm 7)~\mathrm{ps}$. The inset displays the spectrum of the
  transmitted light, recorded with a spectrometer.}
\end{figure}

\begin{figure}[h!]
  \centering
     \includegraphics[width=86mm]{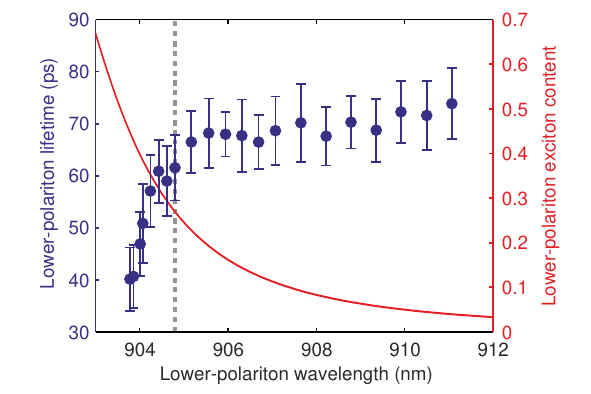}
  \caption{{\bf Polariton lifetimes as a function of cavity detuning
  from the exciton.} Using a pulsed white-light source, the lifetime
  of the fundamental lower polariton was recorded, yielding lifetimes
  up to $74~\mathrm{ps}$ (blue bullets) for almost cavity-like polaritons.
  The red line illustrates the exciton content of the lower polariton that
  was estimated based on PL spectra recorded at the same cavity length. The
  gray dotted line denotes the spectral position of the LP for which the
  lifetime data in the main text were recorded. The cavity length is about
  $14.4~\mu\mathrm{m}$. The error bars correspond to $95~\%$ confidence bounds
  for the fits of the polariton lifetimes.}
\end{figure}

In order to measure polariton lifetimes, we excited the cavity
through the fiber using a pulsed (sub-picosecond at a rate of
$20~\mathrm{MHz}$) white-light source and a band-pass filter,
resulting in a continuous spectrum of about $10~\mathrm{nm}$
bandwidth. The transmitted light was guided out of the cryostat in a
free-space configuration and sent to a spectrometer. We made sure
that only the TEM$_{00}$ LP line appeared in the transmitted
spectrum. We then sent the light to an avalanche photodiode (APD) in
the Geiger mode and measured the ring-down time of the polariton
emission. Figure S3 shows the time-dependent transmission recorded
on the four-QW sample at a cavity length of $28~\mu\mathrm{m}$.
Because of the finite time resolution of the APD (40\,ps), we fit
the ring-down signal (blue trace) by an exponential decay convolved
with the APD response (red trace). The fitted $1/e$ decay time of
the exponential decay corresponds to the polariton lifetime. In
addition to LP lifetimes as a function of the cavity length (see
Figure 3 in the main text), we investigated the lifetime dependence
as a function of cavity-exciton detuning at a cavity length of
$14.4~\mu\mathrm{m}$. At this length, the Rabi splitting is $2 g
\approx 5.8~\mathrm{meV}$. The results are shown in Figure S4. Due
to exciton disorder, the lifetime decreases significantly when the
LP wavelength becomes smaller than $905~\mathrm{nm}$, in agreement
with the line broadening discussed in the previous section. At
$904.8~\mathrm{nm}$, the lifetime is still as high as
$60~\mathrm{ps}$ corresponding to an exciton content of about
$30~\%$. We found this wavelength to be a good compromise between
low disorder scattering and high exciton content. At this
wavelength, the lifetime data of Figure 3 (main text) were recorded.

\pagebreak
%\bibliography{biblio}

\end{document}